\documentclass{article}
\usepackage{ijcai26}

\usepackage{times}
\usepackage{soul}
\usepackage{url}
\usepackage[hidelinks]{hyperref}
\usepackage[utf8]{inputenc}
\usepackage{graphicx}
\usepackage{amsmath}
\usepackage{amsthm}
\usepackage{amssymb}
\usepackage[switch]{lineno}
\usepackage{comment}

\usepackage{balance} 
\usepackage{bm}

\usepackage[ruled,vlined,linesnumbered]{algorithm2e}
\SetKwInOut{Input}{Inputs}
\SetKwInOut{Output}{Outputs}
\SetAlgoNlRelativeSize{-1}      
\SetAlgoInsideSkip{smallskip}   
\SetInd{0.5em}{1.0em}           

\usepackage{booktabs}      
\usepackage{array}         
\usepackage{ragged2e}      
\usepackage{caption}       

\title{Counterfactual Conditional Likelihood Rewards for Multiagent Exploration}

\usepackage{fancyhdr}
\author{
Ayhan Alp Aydeniz$^1$\and
Robert Loftin$^2$\and
Kagan Tumer$^{1}$\\
\affiliations
$^1$Oregon State University\and
$^2$University of Sheffield\\
\emails
\{aydeniza, kagan.tumer\}@oregonstate.edu,
r.loftin@sheffield.ac.uk
}

\begin{document}

\maketitle

\begin{abstract}
    Efficient exploration is critical for multiagent systems to discover coordinated strategies, particularly in open-ended domains such as search and rescue or  planetary surveying. However, when exploration is encouraged only at the individual agent level, it often leads to redundancy, as agents act without awareness of how their teammates are exploring. In this work, we introduce Counterfactual Conditional Likelihood (CCL) rewards, which score each agent’s exploration by isolating its unique contribution to the team exploration. Unlike prior methods that reward agents solely for the novelty of their individual observations, CCL emphasizes observations that are informative with respect to the joint exploration of the team. Experiments in continuous multiagent domains show that CCL rewards accelerate learning for domains with sparse team rewards, where most joint actions yield zero rewards, and are particularly effective in tasks that require tight coordination among agents.
\end{abstract}

\section{Introduction}
Multiagent systems offer great opportunities for scientific information-gathering tasks, ranging from planetary to deep-ocean exploration. In these tasks, success depends on learning to coordinate under sparse feedback, yet such sparsity makes discovering coordinated actions inherently difficult.
A central approach to address sparse rewards in single-agent systems has been to promote exploration through intrinsic rewards~\cite{bellemare2016unifying,tang2017exploration,badia2020never,seo2021state,liu2021behavior,pathak2017curiosity,klyubin2005empowerment,tiomkin2025process}, which encourage agents to diversify experience and uncover informative behaviors even in the absence of external feedback. These methods often struggle in multiagent settings as exploration over joint observations is substantially more complex than in single-agent domains.


One way to induce intrinsic motivation is via exploration based on individual agent observations, which has proven effective in discovering collaborative behaviors~\cite{zamboni2025towards,aydeniz2023novelty,xu2024population}. This decentralized search results in redundant exploration~\cite{zamboni2025towards}, since agents are unaware of each other’s efforts and the collaboration can only be detected by sparse team reward maximization. Addressing this limitation requires reasoning over joint observations, but non-stationarity renders such reasoning intractable: as agents’ policies evolve, the joint observation distribution shifts, destabilizing density-based estimators and embedding networks commonly used in single-agent settings~\cite{seo2021state,badia2020never,liu2021behavior}.

In this work, we introduce counterfactual conditional likelihood (CCL) rewards, enabling effective exploration of the joint state space. By quantifying each agent’s contribution to joint exploration, the proposed reward promotes targeted coverage of the coordinated regions of the state space. In contrast to conventional entropy maximizing approaches~\cite{bellemare2016unifying,badia2020never,seo2021state}, aiming for uniform coverage, CCL rewards prioritize coordinated regions where agent observations are strongly correlated. Crucially, the counterfactual formulation isolates each agent’s contribution, reducing redundant exploration across agents. This strategy is particularly useful when agents must tightly coordinate their actions under sparse rewards.

The key insight in this work is to target coordinated regions of the agents state space rather than uniformly covering each agent's local observation space. We embed each agent’s local observations using a random encoder, avoiding the intractable training of a joint embedding network. For each agent, a counterfactual conditional likelihood is estimated as an intrinsic reward, comparing the likelihood of the agent’s actual observation to that of a counterfactual observation. By rewarding agents for observations that meaningfully affect the joint likelihood, CCL rewards encourage coordinated exploration of the joint state space while reducing redundancy.

The main contribution of this work is a novel reward function for coordinated exploration in multiagent systems under sparse feedback. Unlike existing novelty- or prediction-based signals, our approach explicitly rewards agents for outcomes that are as correlated as needed with their teammates, encouraging the team to discover and consolidate collaborative regions of the state space.

Our experiments in continuous sparse-reward environments show that CCL agents explore more effectively, achieving faster learning and more targeted coverage. In addition, CCL agents obtain substantially higher returns and develop more tightly coordinated behaviors than those trained solely with local observation entropy maximization (OEM). Furthermore, combining local OEM rewards~\cite{seo2021state} with CCL yields even greater performance, uniting decentralized diversity with coordinated joint exploration.


\section{Preliminaries and Related Work}
In this work, we consider cooperative multiagent tasks modeled as an extension of \textit{decentralized partially observable Markov decision processes} (Dec-POMDPs)~\cite{oliehoek2016concise}. It is defined by the tuple $\langle \mathcal{N}, \mathcal{S}, \mathcal{U}, T, \allowbreak \{r_i\}_{i \in \mathcal{N}}, G, O, \gamma \rangle$, where $\mathcal{N}$ is a finite set of agents; $\mathcal{S}$ is the set of environment states; and $\mathcal{U} = \{U^i\}_{i \in \mathcal{N}}$ is the set of joint actions. At each time step $t$, every agent $i \in \mathcal{N}$ selects an action $u_t^i \in U^i$, forming the joint action $\bm{u}_t = \{u_t^i\}_{i \in \mathcal{N}}$. This joint action causes a transition from state $s_t$ to $s_{t+1}$ with probability 
$T(s_t, \bm{u}_t, s_{t+1}) = P(s_{t+1}\mid s_t, \bm{u}_t),$ 
and yields two forms of feedback: an \textit{agent-specific immediate reward} $r_i(s_t, u_t^i)$ for each agent $i$, and a \textit{sparse global reward} $G$ provided to the entire team at the end of the episode. 

In a Dec-POMDP, the true state $s_t$ is not directly observable. Each agent $i$ receives a private observation $o_t^i \in \mathcal{O}^i$ sampled from the observation function $O(\bm{o}_t \mid s_t, \bm{u}_t) = P(\bm{o}_t \mid s_t, \bm{u}_t),$ which specifies a probability distribution over joint observations $\bm{o}_t = \{o_t^1, \dots, o_t^n\}$ given the current state and joint action. These observations may only provide partial and noisy information about the underlying state.

The team objective is to learn a joint policy $\pi(\bm{u}_t \mid \bm{o}_{0:t})$ maximizing the expected discounted return
\begin{equation}
J(\pi) := \mathbb{E}_{\pi}\!\left[\sum_{t=0}^{T-1}\gamma^t \Big(\tfrac{1}{|\mathcal{N}|}\sum_{i \in \mathcal{N}} r_i(s_t, u_t^i)\Big) + \gamma^T G(s_T)\right],
\end{equation}
where $\gamma \in [0,1)$ is the discount factor and $T$ the episode length.

We study sparse reward multiagent tasks requiring agents to coordinate over long horizons to obtain delayed team-level feedback, making exploration and credit assignment challenging. We focus on tightly coupled tasks, where a non-zero reward is obtained only when a minimum number of agents coordinate their actions, defined by a coupling factor. Partial coordination yields no feedback, severely impeding learning.

We address this challenge using task-agnostic intrinsic rewards that promote coordinated exploration without relying on domain-specific shaping.

\subsection{Intrinsic Rewards in Reinforcement Learning}

In RL, intrinsic rewards are task-agnostic signals that promote exploration by encouraging agents to diversify their experiences~\cite{schmidhuber1991possibility,barto2005intrinsic,oudeyer2007intrinsic}. Prior work has proposed a wide range of intrinsic rewards based on novelty, curiosity, entropy maximization, empowerment, or influence~\cite{bellemare2016unifying,tang2017exploration,pathak2017curiosity,klyubin2005empowerment,wang2019influence}. While effective in single-agent domains, these approaches typically assume stationary dynamics and tractable state representations.

In multiagent settings, extending intrinsic rewards becomes substantially more challenging. Joint observation spaces grow combinatorially with the number of agents, rendering density- and entropy-based estimators unreliable, while non-stationarity induced by concurrently learning agents destabilizes learned representations~\cite{badia2020never,seo2021state}. As a result, intrinsic rewards are commonly applied at the agent-level~\cite{zamboni2025towards}, which often leads to redundant exploration and delayed discovery of coordinated behaviors.


Recent work has explored coordinated exploration in multiagent systems through higher-level abstractions or structured state reductions. Hierarchical skill discovery methods learn discrete team- or agent-level skills that promote diverse behaviors without external rewards~\cite{yang2023hierarchical}, while cooperative exploration frameworks guide agents toward shared goals defined in projected or restricted state spaces~\cite{liu2021cooperative}. Formation-aware exploration further reduces the joint exploration space by defining equivalence classes over agent configurations and encouraging diversity over formations~\cite{jo2024fox}. While effective, these approaches rely on explicit abstractions—such as skills, goals, or formations—and do not directly measure how individual agent actions contribute to exploration of the joint state space. In contrast, CCL rewards operate directly on joint observations via counterfactual conditioning, assigning intrinsic rewards based on each agent’s marginal contribution to coordinated exploration.

\section{CCL Agents for Multiagent Exploration}
Departing from prior work, we introduce a reward design that explicitly promotes exploration in the joint observation space. 
Under centralized training with decentralized execution (CTDE)~\cite{foerster2018counterfactual} paradigm, we highlight that the exploration need not be decentralized. Unlike traditional approaches that either treat exploration as an individual process or rely on team rewards discovered only after successful coordination, our approach incentivizes agents to diversify their contributions while avoiding redundancy in the exploration of the state space. While local entropy maximization (decentralized exploration) remains an effective approach for promoting individual exploration, its combination with CCL rewards enables superior exploration, particularly in scenarios where local entropy alone fails to yield coordinated behavior. This synergy leads to more effective, coordinated coverage of the environment.

As the learning framework, we use Multiagent Proximal Policy Optimization (MAPPO)~\cite{kargar2024macrpo,yu2022surprising} with decentralized Long Short-Term Memory (LSTM)~\cite{hochreiter1997long} based actors and a centralized critic. This enables CTDE: agents act only on local observations during rollouts but use the centralized information provided by the critic.

\subsection{Exploration via Entropy Maximization}


State entropy maximization promotes exploration by encouraging diversity in the agent’s state visitation distribution~\cite{liu2021behavior,seo2021state}. Prior work enables entropy estimation in high-dimensional single-agent settings by embedding states into low-dimensional representations, using either learned encoders~\cite{badia2020never} or fixed random encoders~\cite{seo2021state}.



In partially observable multiagent settings, entropy maximization can be formulated at two distinct levels: the individual agent level, promoting diverse local observations, and the team level, which encourages diversity at joint observations.

Local observation entropy maximization (OEM) represents a decentralized form of exploration, as each agent considers only its own observations. Yet, it can still lead to the emergence of collaborative behaviors~\cite{aydeniz2023novelty,xu2024population}. Unless the domain involves complex, high-dimensional observations such as images, these methods typically do not require embedding networks, since they operate directly on each agent’s local observation space. In these settings, the local OEM reward, $r^{Local}_{OEM}$,  that each agent receives is computed based on its own observation history:

\begin{equation}
\label{eq:r_local_OEM}
    r^{Local}_{OEM} = \log(d^{(o^i)}_{\text{knn}} + 1)
\end{equation}

where $d^{(o^i)}_{\text{knn}}$ denotes the distance between the agent’s current observation, $o^i_t$, and its $k$-th nearest neighbor within agent $i$'s independent episodic observation history.

Deriving joint OEM is challenging for two main reasons. First, assigning meaningful credit to each agent for its contribution to the joint entropy is nontrivial. Second, as the joint observation space grows combinatorially with the number of agents, distance-based estimators lose accuracy, and obtaining learned embeddings of the joint observations becomes significantly more difficult due to non-stationarity.

These challenges motivate the need for a formulation that captures each agent’s contribution to joint exploration in a scalable and coordination-aware manner, as developed in our Counterfactual Conditional Likelihood (CCL) framework.

\begin{algorithm}[t]
\small
\caption{Local Observation Entropy Maximization (OEM) Reward for Agent $i$}
\label{alg:oem}
\Input{Initial state $s^0$; initial observation $o_i^0$; horizon $H$; neighbor count $k$.}
\Output{Per-timestep OEM rewards $\{\textit{reward}_t\}_{t=0}^{H-1}$ for agent $i$.}

\BlankLine
\textbf{Init}\; 
$\tau_i^0 \gets \{\,o_i^0\,\}$ \tcp*{Agent-$i$ observation history}

\BlankLine
\For{$t \gets 0$ \KwTo $H-1$}{
  \textbf{Retrieve transition}\;
  \textbf{Observe} $(a_i^t, o^{t}, o_i^{t+1})$ \tcp*{Environment step outcome}
  
  \BlankLine
  \textbf{Compute $k$-nn distance in agent-$i$ history}\;
  $\mathcal{D} \gets \{\, \|\,o_i^{t+1} - o\,\|_2 \;:\; o \in h_i^t \,\}$ \tcp*{Euclidean distances}
  sort $\mathcal{D}$ ascending\;
  $D_{k,t} \gets \text{$k$-th element of } \mathcal{D}$ \tcp*{$k$-NN distance}
  
  \BlankLine
  \textbf{Intrinsic reward}\;
  $r^{Local}_{OEM} \gets \log\!\big(D_{k,t} + 1\big)$ \tcp*{See Eq.~\ref{eq:r_local_OEM}}
  
  \BlankLine
  \textbf{Update history}\;
  \(\tau_i^{t+1} \gets \tau_i^t \cup \{o_i^{t+1}\}\)
  
  \BlankLine
  \textbf{Yield reward} $r^{Local}_{OEM}$ for timestep $t$\;
}
\end{algorithm}

\subsection{Counterfactual Conditional Likelihood Reward Design}
A central challenge in cooperative multiagent exploration is that intrinsic rewards such as prediction error or entropy maximization often fail to scale: agents tend to redundantly explore similar regions of the state space~\cite{zamboni2025towards}, while the true coordinated behaviors needed to solve sparse reward tasks remain undiscovered. To address this, we introduce the \emph{Counterfactual Conditional Likelihood} (CCL) reward, designed to assign each agent an intrinsic reward proportional to its \textit{informational contribution} to the team’s joint observation, capturing how uniquely its behavior influences collective exploration.

The key idea behind the CCL reward is to exploit joint observations effectively.
Rather than estimating entropy directly in this high dimensional and non-stationary space, CCL captures how each agent’s action changes the likelihood of the team’s joint observation. By comparing the actual joint observation with a counterfactual scenario in which the agent remains inactive, the CCL reward quantifies that agent’s unique informational contribution to coordinated exploration. 

We embed each agent’s observations using local random encoders, thereby avoiding to learn intractable joint embeddings for all agents. The concatenation of these local embeddings forms a representation of the team’s joint observation. This structure naturally enables counterfactual reasoning: by substituting an individual agent’s embedding with a counterfactual one, we can estimate how that agent’s behavior influences the likelihood of the overall joint observation.

In computing counterfactuals, we aim to disentangle the influence of an individual agent’s observation from those of its teammates, thereby quantifying its unique contribution to joint exploration. Intuitively, an agent receives a higher reward when its actual embedding is more consistent with the team’s joint behavior than its counterfactual counterpart. By rewarding agents according to their unique informational contribution, CCL discourages redundant exploration and targets the cooperative portion of the state space.

\subsubsection{Formal Definition}
Formally, the CCL reward for agent $i$ is defined as the difference in log-likelihood between the agent’s actual observation and a counterfactual version obtained by holding its observation constant at the previous timestep, conditioned on the observations of all other agents:
\begin{equation}
\label{eq:ccl}
\hat{r}^i_{\text{CCL}} = \log p(o^i_t \mid o^{-i}_t) - \log p(\tilde{o}^i_t \mid o^{-i}_t),
\end{equation}
where $o^{-i}_t = \{ o^1_t, \dots, o^{i-1}_t, o^{i+1}_t, \dots, o^N_t \}$ denotes the observations of all agents except i at time $t$, and $\tilde{o}^i_t$ represents the counterfactual observation of agent $i$, the observation from the previous time step. This formulation measures how much the presence of agent $i$’s actual observation increases the likelihood of the team’s joint observation compared to its counterfactual counterpart. In practice, the likelihoods are estimated using $k$-nn based local density approximations in the embedded joint observation space.

\subsubsection{Estimation in the Embedded Space}
To estimate the likelihoods in Eq.~\ref{eq:ccl}, we first use a random encoder, $\phi$ embedding a local observation, $o^i_t$, into a low dimensional representation, $z^i_t = \phi(o^i_t)$ ($4$ dimensions in our experiments). The encoded representations of all agents are then concatenated to form the team’s joint observation embedding, $z_t = [z^1_t, z^2_t, \allowbreak \dots, z^N_t]$. To compute the \textit{counterfactual} joint embedding for agent $i$, $\tilde z_t^{(i)} = [z^1_t, \dots, z^{i-1}_t, \tilde z^i_t, z^{i+1}_t, \dots, \allowbreak z^N_t]$ , we replace its actual embedding, $z_t^i$ , with its counterfactual embedding $\tilde z^i_t = \phi(o^i_{t-1})$, obtained by encoding the agent’s previous observation, $o^i_{t-1}$.

Within the episodic memory of joint embeddings, $\mathcal{M}$, we compute the distances from both the actual and counterfactual joint embeddings to their respective $k$-th nearest neighbors, denoted by $\epsilon_{\text{act}}$ and $\epsilon_{\text{cfact}}$. To ensure a fair comparison between the two likelihood estimates, we use a shared-radius~\cite{kraskov2004estimating}, where the effective neighborhood radius is defined as $\epsilon_{\text{shared}} = \max(\epsilon_{\text{act}}, \epsilon_{\text{cfact}})$. This shared radius enforces a common local neighborhood for both the actual and counterfactual embeddings, stabilizing the $k$-nn density estimates and reducing sensitivity to local sampling noise~\cite{gao2018demystifying,rodrigues2018combining}.
Using the shared radius, $\epsilon_{\text{shared}}$, we count the number of neighbors within this radius for both the actual and counterfactual embeddings of agent $i$ n the memory $\mathcal{M}$, considering only the embeddings corresponding to agent i. These counts are denoted by $n_{\text{act}}$ and $n_{\text{cfact}}$, respectively.

The conditional log-likelihoods are then approximated using the neighbor counts, $n_{\text{act}}$ and $n_{\text{cfact}}$. Concretely,
\[
\log p(o^i_t \mid o^{-i}t) \;\approx\; \psi\big(n_{\text{act}}+1\big), \quad
\log p(\tilde{o}^i_t \mid o^{-i}t) \;\approx\; \psi\big(n_{\text{cfact}}+1\big),
\]
where $\psi(\cdot)$ is the digamma function. We use the Chebyshev distance metric for all $k$-nn computations for CCL reward, as it provides a consistent measure across spaces of different dimensionalities, the joint embedding space and the agent-specific embedding subspace.

Finally, the CCL reward can be expressed as
\begin{equation}
r^i_{\text{CCL}} = \psi\big(n_{\text{act}}+1\big) - \psi\big(n_{\text{cfact}}+1\big).
\label{eq:ccl-knn}
\end{equation}

The computation procedure is summarized in Algorithm~\ref{alg:ccl}, outlining these steps from encoding to likelihood estimation.

\subsubsection{Reward Shaping and Stabilization}
The scalar reward value computed in the Eq.~\ref{eq:ccl-knn} can be negative. However, to apply the shaping suggested by the work~\cite{aydeniz2023entropy}, we need both stable and positive reward values. To transform the raw counterfactual entropy difference into a stable we apply a Softplus transformation with clamping.
Specifically, the raw value is negated, smoothed via a Softplus activation to ensure non-negativity, and scaled by $\beta=1.0$. then capped to avoid reward explosions. 
Since Softplus is unbounded above, we clamp the output to $\text{cap}=5.0$, thereby preventing reward explosions and ensuring stable gradients during training.

\begin{algorithm}[t]
\small
\caption{Counterfactual Conditional Likelihood (CCL) Reward}
\label{alg:ccl}
\Input{Number of agents $N$; encoder $\phi$ (random, fixed); neighbor count $k$; episodic memory $\mathcal{M}$ of joint embeddings; Chebyshev distance $d_\infty$; Softplus scale $\beta$; clamp cap $\text{cap}$.}
\Output{Per-agent intrinsic rewards $\{r^i_{\mathrm{CCL}}\}_{i=1}^N$.}

\BlankLine
\textbf{Step 1: Encode observations and form joint embedding}\;
\For{$i \gets 1$ \KwTo $N$}{
  $z^i_t \gets \phi(o^i_t)$ \tcp*{Low-dim embedding (4D)}
}
$z_t \gets [z^1_t,\dots,z^N_t]$ \tcp*{Concatenated joint embedding}

\BlankLine
\textbf{Step 2: Build counterfactual joint embedding for each agent}\;
\For{$i \gets 1$ \KwTo $N$}{
  $\tilde z^i_t \gets \phi(o^i_{t-1})$ \tcp*{Hold previous observation}
  $\tilde z^{(i)}_t \gets [z^1_t,\dots,z^{i-1}_t,\tilde z^i_t,z^{i+1}_t,\dots,z^N_t]$\;
  
  \BlankLine
  \textbf{Step 3: KNN radii in joint space (shared-radius)}\;
  $\epsilon_{\mathrm{act}} \gets \text{kth\_radius}(z_t,\mathcal{M},k)$\;
  $\epsilon_{\mathrm{cfact}}  \gets \text{kth\_radius}(\tilde z^{(i)}_t,\mathcal{M},k)$\;
  $\epsilon_{\mathrm{shared}}  \gets \max(\epsilon_{\mathrm{act}},\epsilon_{\mathrm{cfact}})$ \tcp*{KSG-style stabilization~\cite{kraskov2004estimating}}
  
  \BlankLine
  \textbf{Step 4: Counts in agent-$i$ subspace with shared radius}\;
  $n_{\mathrm{act}} \gets \big|\{\,y\!\in\!\mathcal{M}_i : d_\infty(z^i_t,y) < \epsilon_{\mathrm{shared}}\,\}\big|$\;
  $n_{\mathrm{cfact}}  \gets \big|\{\,y\!\in\!\mathcal{M}_i : d_\infty(\tilde z^i_t,y) < \epsilon_{\mathrm{shared}}\,\}\big|$\;
  
  \BlankLine
  \textbf{Step 5: Conditional log-likelihood surrogates}\;
  $\ell_{\mathrm{act}} \gets \psi(n_{\mathrm{act}}+1)$,\quad
  $\ell_{\mathrm{cf}} \gets \psi(n_{\mathrm{cfact}}+1)$ \tcp*{$\psi$: digamma}
  $r^{i}_{\mathrm{CCL}} \gets \ell_{\mathrm{act}} - \ell_{\mathrm{cf}}$\;
  
  \BlankLine
  \textbf{Step 6: Reward shaping (stable, positive)}\;
  $r^{i}_{\text{CCL}} \gets \mathrm{Softplus}(-\,r^{i}_{\mathrm{CCL}})$\;
  $r^{i}_{\mathrm{CCL}} \gets \min(r^{i}_{\text{CCL}},\,\text{cap})$\;
}
\BlankLine
\textbf{Step 7: Update memory}\;
Append $z_t$ to $\mathcal{M}$ \tcp*{Maintain episodic joint-embedding memory}
\end{algorithm}


Then, the per-agent intrinsic reward is further shaped according to the concept of \textit{novelty-seeking rewards}~\cite{aydeniz2023novelty}, applied only in the multi-rover domain experiments. Specifically, the intrinsic reward is multiplied by a saliency value $V(s_t)$ whenever agent $i$ encounters a salient event (\textit{e.g.}, detecting a rarely observed target). The resulting intrinsic signal is then combined with the sparse team reward, $r^{\text{team}}$, which is provided only at the end of an episode:

\begin{equation}
\label{eq:given_rew_0}
    r^i_t \;=\; r^{\text{team}} \,+\, V(s_t)\, r^i_{\text{CCL}}.
\end{equation}

Recent work~\cite{zamboni2025towards} established theoretical bounds highlighting the differences between agent specific entropy maximization and joint entropy maximization, providing trust-region guarantees for both. Building on this, we extend these ideas to continuous and partially observable settings and propose \textit{mixture rewards} that combine local observation entropy~\cite{seo2021state} (computed in Algorithm~\ref{alg:oem}) with joint exploration, CCL rewards. The intuition is that local entropy encourages each agent to diversify its own trajectory, while centralized joint exploration promotes coordinated exploration; combining the two reduces redundant search and enables more efficient coverage of the team’s state space.

We use the local entropy maximizing reward maximization provided in the work~\cite{seo2021state} with local agent observations and derive the mixture rewards:

\begin{figure*}[t]
    \centering
    \includegraphics[width=0.9\linewidth]{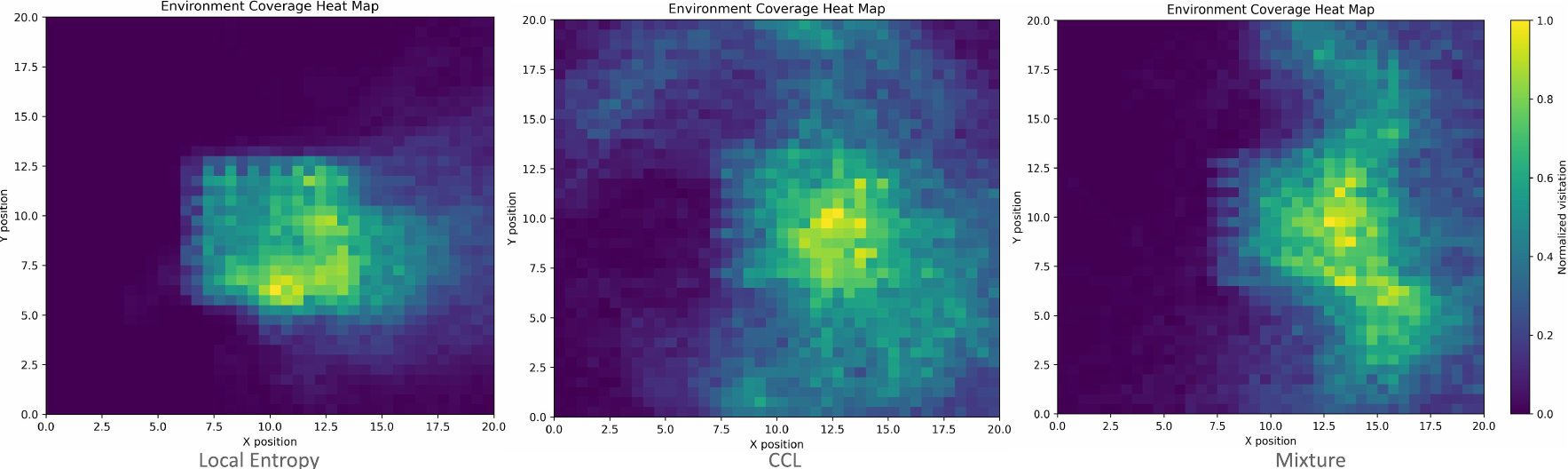}
    \caption{Heat maps of agent trajectories in the multi-rover domain for coupling factor 5 with 2 POIs and 10 agents (Figure~\ref{fig:1cor}). Maps show how agents under different exploration strategies (CCL, Mixture, and Local Entropy) distribute their movements in the environment. CCL encourages more coordinated and complementary coverage of the environment compared to Local Entropy, while the Mixture reward provides intermediate exploration though it achieves earlier convergence to the maximum reward. Each heat map is computed as the average of 20 rollouts collected at iterations 160, 200, 240, 280, 320, 360, and 400.}
    \label{fig:2corHmap}
\end{figure*}

\begin{equation}
\label{eq:given_rew_1}
    r^{\text{mix}}_t \;=\; r^{\text{team}} \,+\, V(s_t)\,\big(r^i_{\text{CCL}} + \alpha \, r^{Local}_{OEM}\big),
\end{equation}
where $\alpha$ (values are provided in Section~\ref{sec:experiments_c4})balances the contribution of local entropy maximization.

To improve stability and reduce sensitivity to neighborhood size, we average the reward estimates across multiple $k$ values (\textit{e.g.}, $k \in \{3, 5, 7\}$). This multi-$k$ averaging provides smoother approximations for both likelihoods and OEM and mitigates outlier effects in sparsely sampled regions.

\begin{table}[htbp]
\centering
\caption{Hyper-parameters used in experiments}
\label{table:hyperparams_c4}
\vspace{-0.2cm}
\label{table:hyper_ccl} 
\catcode`,=\active
\def,{\char`,\allowbreak}
\renewcommand\arraystretch{0.95}
\begin{tabular}{p{2.3cm}<{\raggedright} p{3.0cm} p{2cm}<{\raggedright}}
  \toprule
    \textbf{Component}           & \textbf{Hyperparameter}                  & \textbf{Setting}       \\ 
  \midrule
    MAPPO Core                  & Discount Factor $\gamma$                   & 0.99                   \\
                                & GAE $\lambda$                             & 0.95                   \\
                                & PPO Clip $\epsilon$                       & 0.2                    \\
                                & PPO Epochs                                & 10                     \\
                                & Entropy Coefficient                       & 0.01                   \\
                                & Max Grad. Norm                            & 1.0                    \\
                                & Mini-batch size                   & 32                     \\
                                & Batch Size                                & 37     \\
  \midrule
    Actor / Critic              & Hidden Layer Sizes                        & [128, 128]             \\
                                & Recurrent Layers                          & 1                      \\
                                & Activation Func.                          & ReLU                   \\
                                & Trainable Std. Dev.                       & True                   \\
                                & Init Log Std. Dev.                        & 0.01                   \\
                                & Actor Learning Rate                       & 1e-3                   \\
                                & Critic Learning Rate                      & 1e-3                    \\
                                & Optimizer                                 & Adam~\cite{kingma2014adam} \\
  \midrule
    Training Setup              & Rollout Steps                             & 1200                   \\
  \midrule
    CCL Reward                  & $k$ (nearest neighbors)                   &  \{3,5,7\}             \\
                                & cap (max clamp for stability)             & 5.0                    \\
                                & embedded obs. dim.                                   & 4                    \\
  \midrule                             
   Local Entropy                & k                                         & \{3,5,7\}                    \\

    \midrule
    Episode Length &
Rover domain: 50 \\
& PEnvs: 80 \\
  \bottomrule
\end{tabular}
\end{table}

\section{Experiments}
\label{sec:experiments_c4}
Our experiments are designed to answer the following questions:  
(i) \textit{Does the proposed Counterfactual Conditional Likelihood (CCL) reward improve exploration efficiency in sparse reward multiagent tasks compared to local observation entropy maximization}?  
(ii) \textit{How does CCL affect coordination quality, i.e., the ability of agents to discover complementary behaviors rather than redundant ones}?  
(iii) \textit{What is the effect of combining CCL with local entropy maximization via mixture rewards}?  
(iv) \textit{How robust is CCL to variations in task complexity, such as the number of agents, task difficulty, and sparsity in the environment}?

All hyperparameters are listed in Table~\ref{table:hyperparams_c4}. Results are averaged over 10 independent runs with random seeds. The $x$-axis denotes the number of \emph{policy update iterations}, each consisting of data collection (1200 rollout steps) followed by a PPO update. Policies are evaluated after each iteration over 20 episodes, and report the average team reward; shaded regions indicate one standard deviation across the evaluation episodes and independent seeds.

Agents are trained with MAPPO, with CCL, mixture, and local observation entropy maximizing rewards incorporated. Performance metrics include average team reward, coverage of the environment in both particle environments~\cite{penv_original} in partially observable and sparse reward settings, and multi-rover domain~\cite{agogino2004unifying} in tightly coupled and sparse reward settings.

All agents share an LSTM-based architecture and a fixed random encoder $\phi$ used for CCL reward computation. The encoder is a three-layer MLP (64 units each, \texttt{LayerNorm}, \texttt{SiLU} activations), following the paper~\cite{seo2021state}.

\subsection{Environments}
We evaluate our approach in several benchmark environments. Below we provide a brief overview of each domain, while referring to the original sources for full details on observation spaces, action sets, and reward formulations~\cite{majumdar2020evolutionary,penv_original}.

\subsubsection{Multi-Rover Exploration}
\label{sec:rover_domain}
The multi-rover domain~\cite{agogino2004unifying} is a continuous environment with sparse rewards, where the global team reward is only assigned at the end of each episode. Multiple agents (\textit{rovers}) must learn cooperative strategies in order to simultaneously observe points of interest (POIs). Each POI requires a minimum number of rovers, specified by a \textbf{coupling factor}, for it to be successfully observed. Higher coupling factors increase the difficulty of coordination. Solving this task requires rovers to execute long sequences of coordinated actions, which makes the domain particularly challenging under sparse rewards. In this domain, as suggested by the work~\cite{aydeniz2023novelty}, we use the value of the POI as the value of saliency $V(s)$ while constructing the rewards given in Eqs.~\ref{eq:given_rew_0}, and ~\ref{eq:given_rew_1}. Important to note that this value is only incorporated when agents enter the observation radius of the POI, making POI discovery particularly challenging. The global reward is only provided at the end of an episode.

\begin{figure}[htbp]
    \centering
    \includegraphics[width=1.0\linewidth]{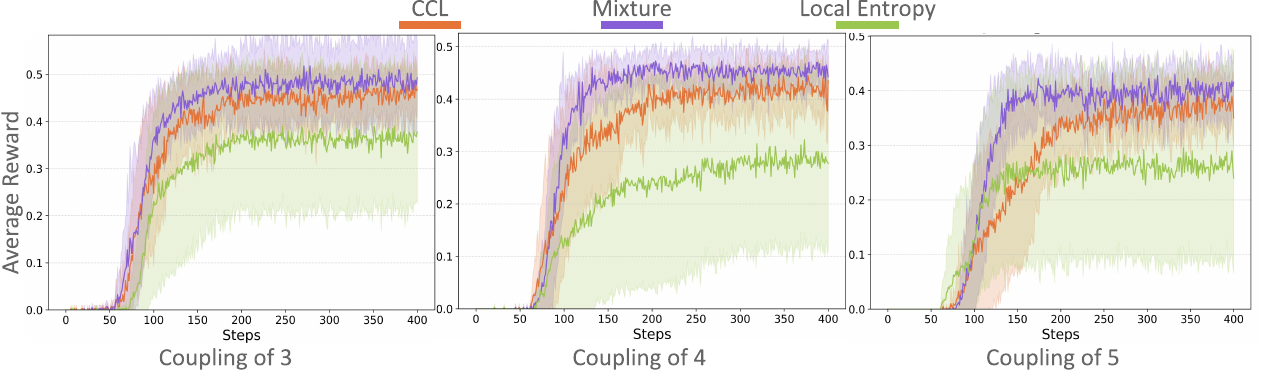}
    \caption{Comparison of exploration strategies in the multi-rover domain across different coupling factors 3, 4, and 5, with teams of 6, 8, and 10 agents, respectively. The environment has two distantly placed POIs. Results show that CCL improves coordinated behaviors and achieve higher performance than the local entropy agents. The mixture reward enhances performance under higher coupling requirements.}
    \label{fig:2cor}
\end{figure}

\subsubsection{Particle Environments}
We also consider three multiagent particle environments~\cite{penv_original} in partially observable and sparse rewards settings. Each scenario includes three cooperative agents (\textit{good}) and a single adversary. The adversary is trained using PPO~\cite{ppo}, with the same policy architecture as the cooperative agents. The episodic team reward is defined as the sum of cooperative agents’ rewards. The tasks are as follows:
\begin{itemize}
    \item \textit{Physical deception}: Cooperative agents and an adversary compete to reach a shared landmark. The cooperative agents’ reward combines their distance to the landmark with a penalty based on the adversary’s distance, incentivizing coordination to block the adversary.
    \item \textit{Keep away}: Cooperative agents must reach a target landmark (randomly selected from two) and are rewarded based on their distance to it. The adversary aims to disrupt them by pushing them away from the chosen target.
    \item \textit{Predator–prey}: Here, landmarks serve as obstacles. Cooperative agents, which move more slowly, must capture a faster adversary. The cooperative team receives a positive reward when contact is made with the adversary, while the adversary is penalized for being caught.
\end{itemize}

\begin{figure*}[t]
    \centering
    \includegraphics[width=0.9\linewidth]{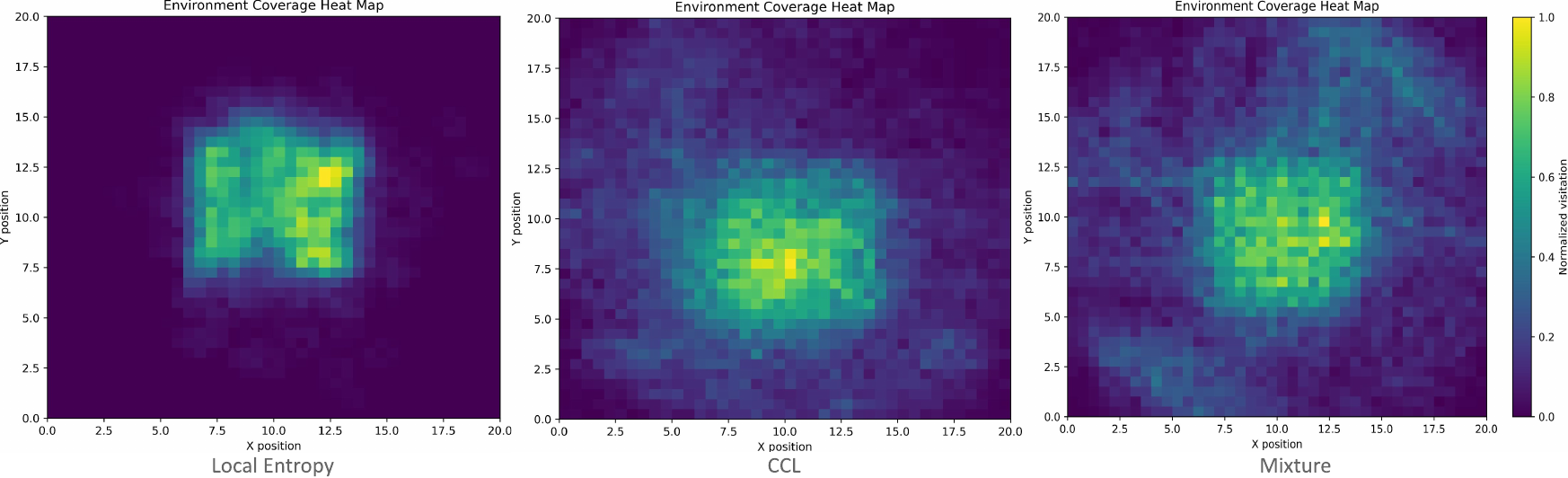}
    \caption{Heat maps showing agent trajectories in the multi-rover environment with 5 agents, a coupling factor of 5, and a single POI. Results are averaged over 20 rollouts at training iterations 160, 200, 240, 280, 320, 360, and 400. The plots illustrate how CCL fosters coordinated exploration toward the POI, while local entropy leads to failed expansion and redundant visits; mixture reward produce slightly better expansion then the CCL rewards.}
    \label{fig:1corHmap}
\end{figure*}

In these environments, we do not provide any heuristic for the saliency ($V(s)$ is not defined in Eqs.~\ref{eq:given_rew_0}, and ~\ref{eq:given_rew_1}.). The agents solely receive the intrinsic rewards and the global team reward is only provided at the end of an episode.


\section{Results}

Our experiments evaluate the proposed Counterfactual Conditional Likelihood (CCL) reward across cooperative multiagent tasks. We present results in line with the four guiding questions posed in Section~\ref{sec:experiments_c4}.

\begin{figure}[htbp]
    \centering
    \includegraphics[width=1.0\linewidth]{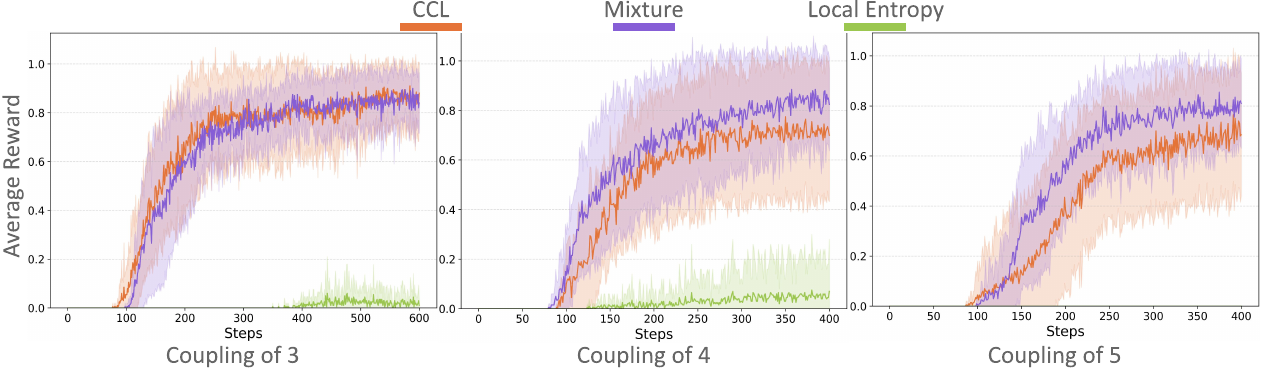}
    \caption{Comparison of exploration strategies in the multi-rover domain across different coupling factors 3, 4, 5, and 6, with teams of 3, 4, 5, and 6 agents, respectively. The environment has only 1 POI. This setting presents a bigger challenge, because there is only a single distantly placed POI. It is difficult to discover. Plots show the importance of joint exploration. CCL improves coordinated behaviors and achieve high reward, while the local entropy maximizing agents fail. Mixture rewards provide additional gains; however, their impact diminishes compared to Figure~\ref{fig:2cor}.}
    \label{fig:1cor}
\end{figure}

\paragraph{Exploration Efficiency.}  
Figures~\ref{fig:2cor} and~\ref{fig:1cor} demonstrate that CCL substantially improves exploration in sparse reward multi-rover domain configurations. In settings with two POIs, the task becomes less sparse since agents have more opportunities to increase the team reward above zero. In this case, agents trained with local entropy maximization achieve moderate performance, whereas CCL consistently outperforms across all coupling factors. Importantly, we also increase the number of agents in the system. The gains are more pronounced when the environment has a single POI (Figure~\ref{fig:1cor}), a particularly challenging setting where discovery requires tightly synchronized behaviors. There is only a single POI, making the discovery of behaviors achieving non-zero reward much more difficult (The reward can either be 0 or 1). The agents trained with local entropy fail to observe a POI. By contrast, CCL consistently drives agents toward complementary exploration patterns, uncovering coordinated solutions. We can see the benefits of coordinated joint exploration.

To assess whether CCL promotes complementary behaviors, we examine agent trajectories directly. Heat maps in Figure~\ref{fig:2corHmap} show that local entropy maximization leads to poor exploration. In As shown in Figure~\ref{fig:1corHmap}, local entropy agents fail in the single-POI environment because their exploration is uncoordinated, with agents clustering in the same regions rather than jointly covering the environment. CCL reduces such redundancy by rewarding unique contributions to joint exploration, resulting in broader, more balanced coverage. This coordinated exploration underpins the higher team rewards observed in Figures~\ref{fig:2cor} and~\ref{fig:1cor}.

\begin{figure}[htbp]
    \centering
    \includegraphics[width=1.0\linewidth]{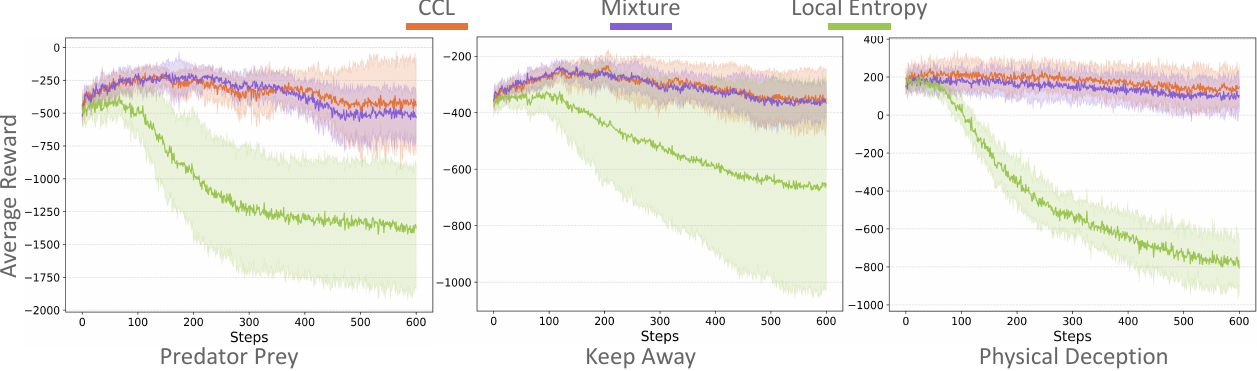}
    \caption{Comparison of exploration strategies in particle environments (Predator–Prey, Keep Away, and Physical Deception). For all scenarios, there are 3 good agents and 1 adversarial agent. Results show that CCL improves coordination and overall team reward compared to local entropy maximization, which struggles to uncover cooperative strategies. Mixture rewards provide further improvements, though the relative benefit varies across tasks.}
    \label{fig:penv}
\end{figure}

\paragraph{Mixture Rewards.}  
Combining CCL with local entropy maximization provides additional benefits, particularly in the early stages of training (Figures~\ref{fig:2cor} and~\ref{fig:penv}). Mixture rewards achieve faster convergence and higher peak performance in some settings. However, their relative advantage diminishes in harder tasks such as the single-POI rover domain (Figure~\ref{fig:1cor}), where joint coordination is indispensable. These findings suggest that mixture rewards offer a useful trade-off between local diversity and joint exploration, but are less effective when coordination demands dominate. 

We can also analyze their impact through the $\alpha$ parameter used in the Eq.~\ref{eq:given_rew_1}. We conducted experiments varying $\alpha$ over ${0.1, 0.3, 0.5, \allowbreak 0.7, \allowbreak 1.0}$. Among these, $\alpha = 0.5$ consistently produced the highest average reward for the mixture reward, particularly in settings where local entropy agents achieved moderate performance. However, in scenarios where the local entropy maximizing agents fail, the alpha value of $0.1$ produced the highest performance.

Overall, the findings indicate that CCL and local entropy agents specialize in exploring different regions of the state space, making their combination particularly effective for coordinated exploration in environments with denser tasks.

Finally, Figure~\ref{fig:penv} evaluates CCL in diverse particle environments with adversarial agents. Across Predator–Prey, Keep Away, and Physical Deception, CCL consistently outperforms local entropy. Together with results from the rover domain, these findings indicate that CCL generalizes robustly across domains, scaling effectively with task complexity, number of agents, and reward sparsity.


\section{Conclusion}

In this work, we introduced the \emph{Counterfactual Conditional Likelihood} (CCL) reward, a novel intrinsic reward for coordinated exploration in cooperative multiagent systems. Unlike standard entropy-based approaches relying on local observations or unstable joint embeddings, CCL uses counterfactuals to isolate each agent’s contribution to joint exploration.  

Across continuous multi-rover domains and particle environments, we showed that CCL improves exploration efficiency, reduces uncoordinated behavior, and enables agents to discover complementary strategies in sparse reward tasks. CCL consistencly outperforms local entropy maximization across varying task sparsity and complexity, while mixture rewards provide a practical compromise that accelerates learning in simpler settings. Importantly, the benefits of CCL extend beyond a single environment class, generalizing to adversarial multiagent scenarios where coordination is critical.  

Overall, CCL provides a principled approach to bridging local novelty incentives and global coordination in cooperative MARL. As team size grows, the accuracy of particle-based density estimators degrades, which may reduce CCL’s effectiveness; addressing this limitation and integrating model-based exploration strategies are important directions for future work~\cite{klyubin2005empowerment,jaques2019social}.

\bibliographystyle{named}
\bibliography{ijcai26}

\end{document}